\documentclass[aps,prl,twocolumn,showpacs,amsmath,amssymb, nofootinbib]{revtex4-1}
\usepackage{epsfig}
\usepackage{graphicx}
\usepackage{dcolumn}
\usepackage{bm}
\usepackage{ltablex,booktabs}
\usepackage{overpic}
\usepackage{subfigure}
\usepackage{float}
\usepackage{color}
\usepackage{amsmath}
\usepackage{mathcomp}
\usepackage{mathrsfs}
\usepackage{multirow}
\usepackage{rotating}
\usepackage{amssymb}
\usepackage{gensymb}
\usepackage{amsmath}
\usepackage{tabularx}
\usepackage{tikz}
\usepackage{tikz-feynman}
\usepackage[bookmarksnumbered, pdfstartview=FitH,colorlinks,urlcolor=blue, citecolor=blue,linkcolor=blue] {hyperref}

\begin{document}
\normalsize
\parskip=5pt plus 1pt minus 1pt


\title{Observation of $D^{+}\to K_{S}^{0}a_{0}(980)^{+}$ in the amplitude analysis of $D^{+} \to K_{S}^{0}\pi^+\eta$}
\author{
\begin{small}
\begin{center}
M.~Ablikim$^{1}$, M.~N.~Achasov$^{4,b}$, P.~Adlarson$^{75}$, X.~C.~Ai$^{81}$, R.~Aliberti$^{35}$, A.~Amoroso$^{74A,74C}$, M.~R.~An$^{39}$, Q.~An$^{71,58}$, Y.~Bai$^{57}$, O.~Bakina$^{36}$
% ...
\\
\vspace{0.2cm}
(BESIII Collaboration)\\
\vspace{0.2cm} {\it
$^{1}$ Institute of High Energy Physics, Beijing 100049, People's Republic of China\\
% ...
$^{4}$ Budker Institute of Nuclear Physics SB RAS (BINP), Novosibirsk 630090, Russia\\
% ...
$^{35}$ Johannes Gutenberg University of Mainz, Johann-Joachim-Becher-Weg 45, D-55099 Mainz, Germany\\
$^{36}$ Joint Institute for Nuclear Research, 141980 Dubna, Moscow region, Russia\\
% ...
$^{39}$ Liaoning Normal University, Dalian 116029, People's Republic of China\\
% ...
$^{57}$ Southeast University, Nanjing 211100, People's Republic of China\\
$^{58}$ State Key Laboratory of Particle Detection and Electronics, Beijing 100049, Hefei 230026, People's Republic of China\\
% ...
$^{71}$ University of Science and Technology of China, Hefei 230026, People's Republic of China\\
% ...
$^{74}$ University of Turin and INFN, (A)University of Turin, I-10125, Turin, Italy; (B)University of Eastern Piedmont, I-15121, Alessandria, Italy; (C)INFN, I-10125, Turin, Italy\\
$^{75}$ Uppsala University, Box 516, SE-75120 Uppsala, Sweden\\
% ...
$^{81}$ Zhengzhou University, Zhengzhou 450001, People's Republic of China\\
\vspace{0.2cm}
% ...
$^{b}$ Also at the Novosibirsk State University, Novosibirsk, 630090, Russia\\
% ...
}
\end{center}
\vspace{0.4cm}
\end{small}
}
\noaffiliation{}

\date{\today}
  
\begin{abstract}
  We perform for the first time an amplitude analysis of the decay
  $D^{+}\to K_{S}^{0}\pi^+\eta$ and report the observation of the decay
  $D^{+}\to K_{S}^{0}a_{0}(980)^{+}$ using 2.93~fb$^{-1}$ of $e^+e^-$
  collision data taken at a center-of-mass energy of 3.773~GeV with the BESIII
  detector. As the only $W$-annihilation-free decay among $D$ to
  $a_{0}(980)$-pseudoscalar $D^{+}\to K_{S}^{0}a_{0}(980)^{+}$ is the ideal
  decay in extracting the contributions of the $W$-emission amplitudes
  involving $a_{0}(980)$ and to study the final-state interactions. The
  absolute branching fraction of $D^{+}\to K_{S}^{0}\pi^+\eta$ is measured to
  be $(1.27\pm0.04_{\rm stat.}\pm0.03_{\rm syst.})\%$. The branching
  fractions of intermediate processes $D^{+}\to K_{S}^{0}a_{0}(980)^{+}$ with
  $a_{0}(980)^{+}\to \pi^+\eta$ and $D^{+}\to \pi^+ \bar{K}_0^*(1430)^0$ with
  $\bar{K}_0^*(1430)^0\to K_{S}^{0}\eta$ are measured to be
  $(1.33\pm0.05_{\rm stat.}\pm0.04_{\rm syst.})\%$ and
  $(0.14\pm0.03_{\rm stat.}\pm0.01_{\rm syst.})\%$, respectively.
\end{abstract}
\maketitle

Perturbative quantum chromodynamics~(QCD) approaches, such as QCD factorization
and soft-collinear effective theory, have well explained physics of nonleptonic
$b$-hadron
decays~\cite{Beneke:2001ev,Leibovich:2003tw,Han:2022srw,Hsiao:2022tfj}.
However, the charm quark mass is located between the perturbative and
nonperturbative QCD regions, making neither of those approaches applicable.
As a result, an accurate theoretical description of the underlying mechanism
for exclusive hadronic decays of charmed mesons is still not available. A
model-independent analysis was then proposed in the so-called diagrammatic
approach to phenomenologically describe charmed meson
decays~\cite{Chau:1986jb}. The diagrammatic approach represents the
$W$-emission and weak-annihilation~($W$-exchange or $W$-annihilation)
amplitudes as topological quark-graph diagrams based on SU(3) flavor symmetry.
With necessary experimental inputs, it enables us to extract the contribution
and study the relative importance of each amplitude.

On the other hand, great progress has been achieved by a series of amplitude
analyses on the hadronic charmed meson
decays~\cite{PDG, BESIII:2022zx, BESIII:2022lzh, BESIII:2022gpt, BESIII:2022st, BESIII:2022twh, BESIII:2022kbq, BESIII:2021cdy, BESIII:2019kbq, BESIII:2019ly, BESIII:2017ly}.
According to these studies, $D$ meson decays are dominated by quasi-two-body
processes, such as $D\to PP$, $VP$, $SP$, $AP$, and $TP$, where $P$, $V$, $S$,
$A$, and $T$ denote pseudoscalar, vector, scalar, axial-vector, and tensor
mesons, respectively. These enriching experimental results allow the diagrammatic
approach to be applied quite successfully in $D\to PP$ and $D\to VP$
decays~\cite{Cheng:VP2019, Cheng:VP2010}. However, in the sector of $D\to SP$,
it appears that the current experimental measurements are still
insufficient~\cite{Cheng:2002ai, Cheng:2010vk, Cheng:2022vbw, Hussain:1986xw, Katoch:1994ux, Fajfer:1995hp, Buccella:1996uy, El-Bennich:2008rkp, Boito:2008zk, Dedonder:2014xpa, Xie:2014tma, Dedonder:2021dmb},
and the decay $D^+ \to K_S^0a_0(980)^+$ is the most urgently needed and ideal
decay to test the validity of the diagrammatic approach.

\begin{figure}[htbp]
  \centering
  \includegraphics[width=0.235\textwidth]{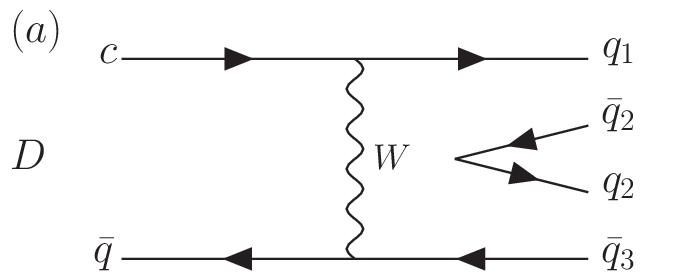}
  \includegraphics[width=0.235\textwidth]{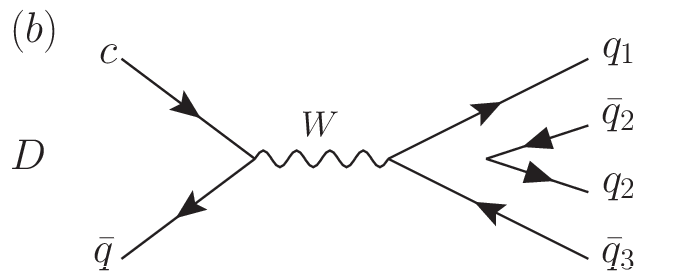}
  \caption{(a) The $W$-exchange and (b) the $W$-annihilation diagrams for $D$
    decays. For the $D^+$ meson, the $W$-exchange mechanism is simply absent.
    The $W$-annihilation mechanism cannot generate the hadronic mode with a
    $\bar{K}^0$ in the final state.
  }
\label{feynmen}
\end{figure}

\begin{figure}[htbp]
  \centering
  \includegraphics[width=0.235\textwidth]{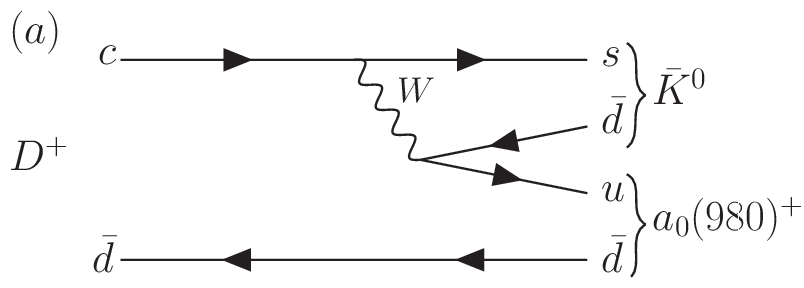}
  \includegraphics[width=0.235\textwidth]{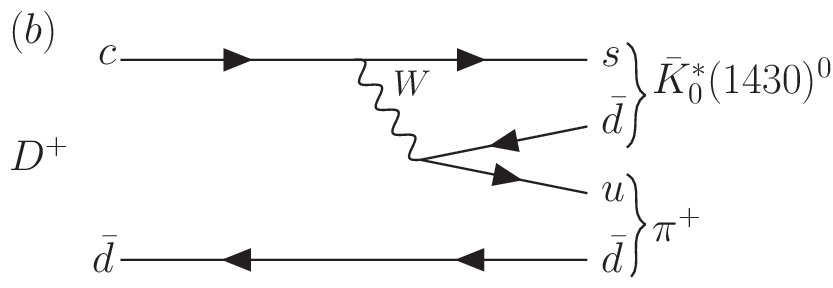}
  \includegraphics[width=0.235\textwidth]{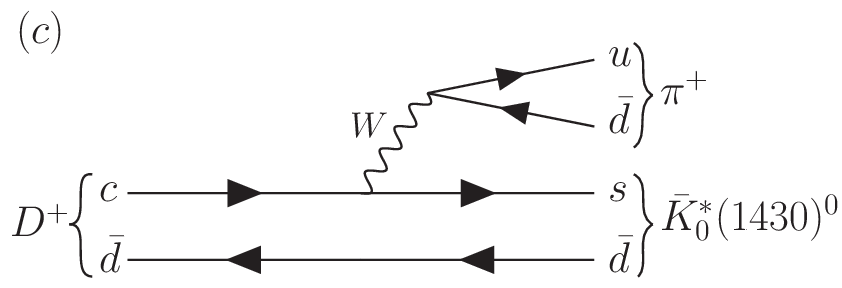}
  \includegraphics[width=0.235\textwidth]{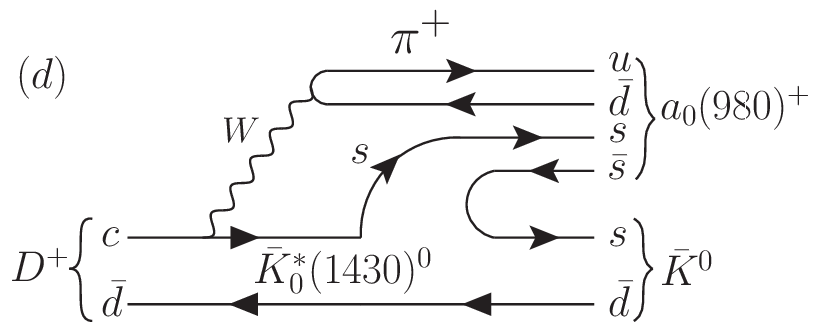}
  \caption{(a) The internal $W$-emission diagram for the decay 
    $D^+\to\bar{K}^0a_0(980)^+$ by assuming $a_0(980)^+$ is a two-quark state.
    (b) The internal and (c) the external $W$-emission diagrams for the decay
    $D^+\to\bar{K}_0^*(1430)^0\pi^+$. (d) An example diagram of
    $D^+\to\bar{K}^0a_0(980)^+$ via rescattering in the final state of
    $D^+\to\bar{K}_0^*(1430)^0\pi^+$ under the assumption that $a_0(980)^+$ is
    a tetraquark state.
  }
\label{feynmen2}
\end{figure}

A large discrepancy between experimental results and theoretical predictions
of branching fractions~(BFs) for many $D^{+,0}\to a_0(980)P$ decays have been
found~\cite{Cheng:2002ai, Cheng:2010vk, Cheng:2022vbw}. The main reason could
be ascribed to the contribution of the weak-annihilation amplitudes in $D$
decays, which are hard to estimate accurately. Among $D^{+,0} \to a_0(980) P$,
$D^+ \to K_S^0a_0(980)^+$ is the only decay free of weak-annihilation
contributions, as depicted in Fig.~\ref{feynmen}, and mainly involves the
internal $W$-emission in Fig.~\ref{feynmen2}(a), while its BF is not
theoretically predicted. Without the interference from weak annihilation, the
study of $D^+ \to K_S^0a_0(980)^+$ will serve as a key experimental input and
provide the most sensitive constraint to the contributions and phases of the
internal $W$-emission amplitudes involving $a_0(980)$ in the diagrammatic
approach method~\cite{Cheng:2002ai, Cheng:2010vk, Cheng:2022vbw}. It is worth
noting that the external $W$-emission amplitude for this decay is naively
expected to be rather suppressed compared to the internal one, due to
$G$-parity violation~\cite{BCKa0, BCKa02}. In addition, the light scalar
particle $a_0(980)$ is commonly considered as a candidate for exotic states,
which are states other than typical quark-antiquark mesons, such as states for
tetraquarks, $K\bar K$ bound states, and other possible states. The production
of these exotic states essentially involves final-state interactions, such as
quark exchange, resonance formation,
etc.~\cite{BCKa03, BCKa0, BCKa02, Zhang:2022xpf, Achasov:2021dvt, Achasov:2017edm}.
For example, the production of the tetraquark $a_0(980)^+$ state in $D^+$
decays can occur as a result of the fact that the seed tetraquark fluctuations
$u\bar{d}\to\pi^+\eta$, $\pi^+\eta^\prime$, $K^+\bar{K}^0$ are dressed by
strong resonance interactions in the final state~\cite{Achasov:2021dvt}.
Figure~\ref{feynmen2}(d) also illustrates an example of the production of the
tetraquark $a_0(980)^+$ state due to rescattering in the final state. Studying
$D^+ \to K_S^0a_0(980)^+$ can experimentally constrain the contribution from
these effects, which helps to pin down the nature of $a_0(980)$.


Moreover, precision measurements of absolute hadronic $D$ meson BFs are
essential for both charm and beauty physics. The $D^+\to K_S^0\pi^+\eta$ decay,
dominated by the Cabibbo-favored process, has a large BF of the order of
$10^{-2}$. This decay contains other possible intermediate amplitudes, such as
$\bar{K}_0^*(1430)^0\pi^+$, as depicted in
Figs.~\ref{feynmen2}(b)~and~\ref{feynmen2}(c). Studying the relative
contribution of the intermediate resonances can not only benefit the
understanding of the strong interaction at low energies but also determine
these missing $D^+$ decay modes. 

The BESIII detector collected 2.93 fb$^{-1}$ of $e^+e^-$ collision data in 2010 
and 2011 at $\sqrt{s}=3.773$~GeV~\cite{Ablikim:2013ntc}, which corresponds to 
the mass of the $\psi(3770)$ resonance. The $\psi(3770)$ decays predominantly 
to $D^0\bar D^{0}$ or $D^+D^-$ without any additional hadrons. The excellent 
tracking, precision calorimetry, and the large $D\bar{D}$ threshold data sample 
provide an unprecedented opportunity to study the charmed meson decays.
Based on this dataset, we present the first amplitude
analysis of the decay $D^+\to K_S^0\pi^+\eta$~\cite{PRL124.241803} and report
the observation of the decay $D^+ \to K_S^0 a_0(980)$. Charge-conjugate states
are implied throughout this Letter.

The BESIII detector~\cite{Ablikim:2009aa} records the final-state particles of
symmetric $e^+e^-$ collisions provided by the BEPCII storage
ring~\cite{Yu:IPAC2016-TUYA01} in the center-of-mass energy range from 2.00 to
4.95~GeV, with a peak luminosity of
$1 \times 10^{33}\;\text{cm}^{-2}\text{s}^{-1}$ achieved at
$\sqrt{s} = 3.77\;\text{GeV}$. The cylindrical core 
of the BESIII detector covers 93\% of the full solid angle and consists of a 
helium-based multilayer drift chamber, a plastic scintillator 
time-of-flight system, and a CsI(Tl) electromagnetic calorimeter,
which are all enclosed in a superconducting solenoidal magnet providing a
1.0~T magnetic field.

Simulated data samples produced with a {\sc geant4}-based~\cite{geant4} Monte 
Carlo (MC) toolkit, which includes the geometric description~\cite{detvis} of
the BESIII detector and the detector response, are used to determine detection 
efficiencies and to estimate backgrounds. The simulation models the beam energy 
spread and initial state radiation (ISR) in the $e^+e^-$ annihilations with the 
generator {\sc kkmc}~\cite{ref:kkmc}. The inclusive MC sample includes the 
production of $D\bar{D}$ pairs (including quantum coherence for the neutral $D$ 
channels), the non-$D\bar{D}$ decays of the $\psi(3770)$, the ISR production of 
the $J/\psi$ and $\psi(3686)$ states, and the continuum processes incorporated 
in {\sc kkmc}. All particle decays are modelled with 
{\sc evtgen}~\cite{ref:evtgen} using BFs either taken from 
the Particle Data Group~(PDG)~\cite{PDG}, when available, or otherwise
estimated with {\sc lundcharm}~\cite{ref:lundcharm}. Final-state radiation from 
charged final-state particles is incorporated using {\sc photos}~\cite{photos}.

By fully reconstructing the $D^+D^-$ meson pairs, a double-tag (DT) method
provides samples with high purity to perform amplitude analyses and
measurements of absolute BFs of the hadronic $D^+$ meson decays. The DT
candidates are required to be the $D^{+}$ meson decaying to the signal mode
$D^{+} \to K_{S}^{0}\pi^{+}\eta$ and the $D^{-}$ meson decaying to six tag
modes: $K^{+}\pi^{-}\pi^{-}$, $K_{S}^{0}\pi^{-}$, $K^{+}\pi^{-}\pi^{-}\pi^{0}$,
$K_{S}^{0}\pi^{-}\pi^{0}$, $K_{S}^{0}\pi^{-}\pi^{-}\pi^{+}$, or
$K^{+}K^{-}\pi^{-}$. The selection criteria for the final-state particles are
the same as in Ref.~\cite{PRL124.241803}.

The $D^\pm$ mesons are selected using two variables, the energy difference
$\Delta E = E_{D} - E_{\rm b}$ and the beam-constrained mass
$M_{\rm BC} = \sqrt{E^{2}_{\rm b}-|\vec{p}_{D}|^{2}}$, where $E_{\rm b}$ is the
beam energy and $\vec{p}_{D}$ and $E_{D}$ are the momentum and the energy of
the $D^\pm$ candidate in the $e^+e^-$ rest frame, respectively. The $D^-$ meson
is reconstructed first through the six tag modes. In case of multiple
candidates, the one with the minimum $|\Delta{E}|$ is chosen. Once a tag is
identified, the signal decay $D^{+} \to K_S^0\pi^{+}\eta$ is searched for at
the recoiling side and the best signal candidate with the minimum $|\Delta{E}|$
is selected. All $D^{\pm}$ candidates are required to satisfy
$1.865<M_{\rm BC}<1.875$~GeV, and $-0.055<\Delta{E}<0.040$~GeV for the tag
modes containing $\pi^0$ in the final state, $-0.025<\Delta{E}<0.025$~GeV for
the others, and $-0.020<\Delta{E}<0.020$~GeV for the signal side. In addition,
the energy of the largest unused photons is required to be less than 0.23~GeV.
There are 1113 DT events obtained for the amplitude analysis with a signal
purity of $(98.2\pm0.4)\%$, which is determined from a two-dimensional
unbinned maximum-likelihood fit to the distribution of $M_{\rm BC}$ of the tag
$D^-$ versus that of the signal $D^+$.
(Details of the fit are introduced in Ref.~\cite{PRL124.241803}.)

\begin{figure}[htbp]
  \centering
  \includegraphics[width=0.235\textwidth]{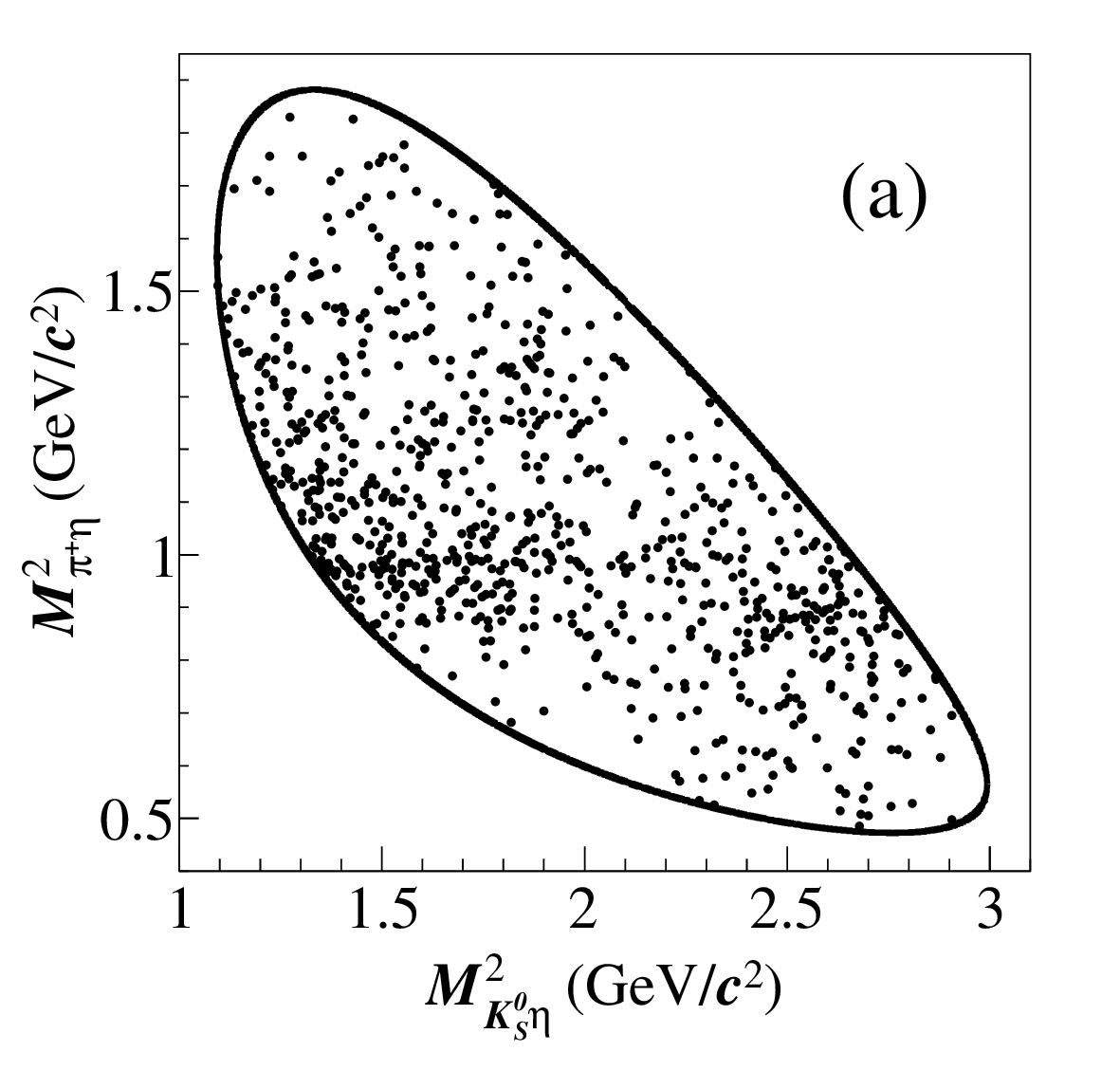}
  \includegraphics[width=0.235\textwidth]{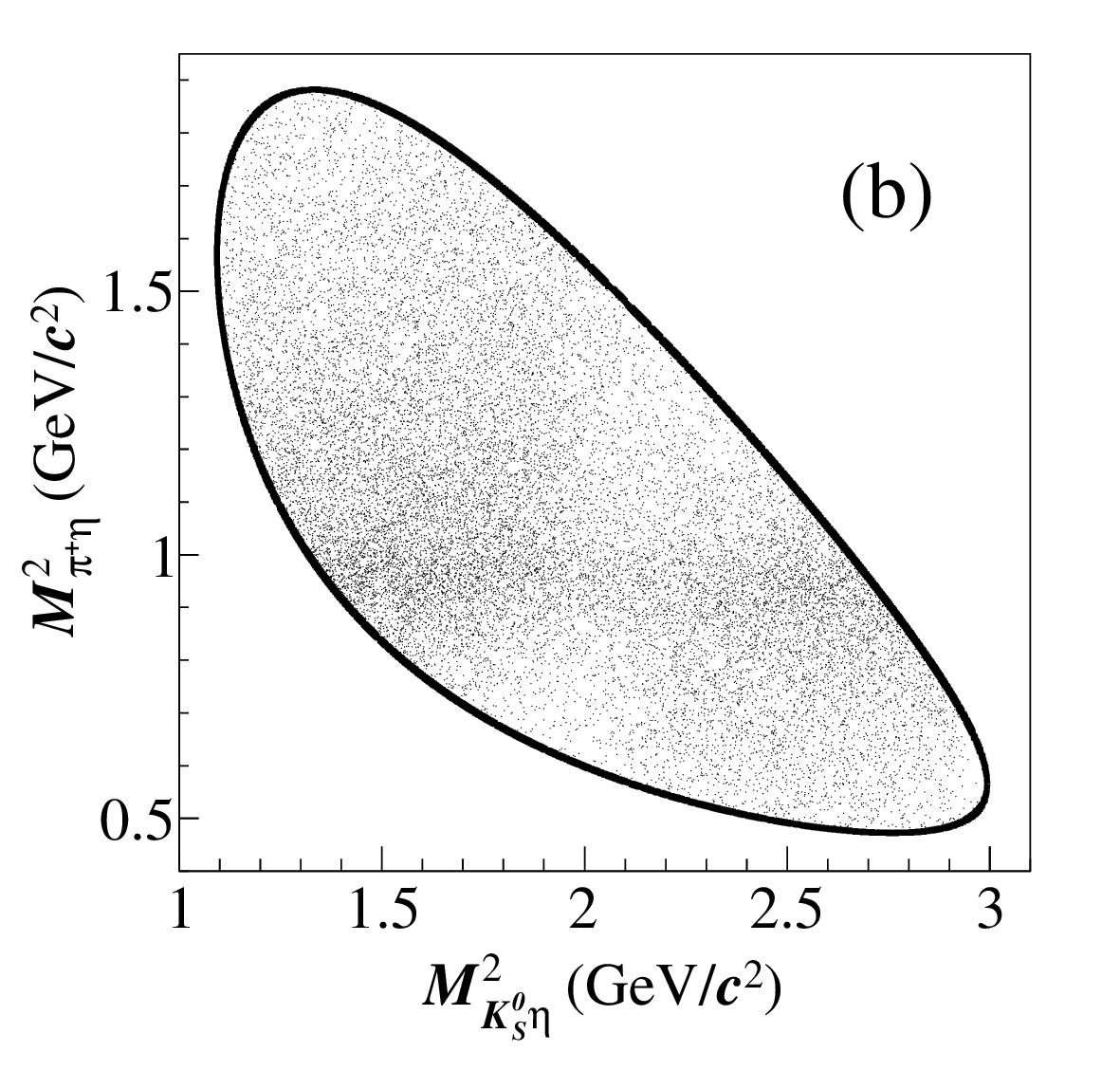}
  \caption{The Dalitz plots of $M^{2}_{\pi^+\eta}$ versus $M^{2}_{K_S^0\eta}$
    of the $D^+\to K^0_S\pi^+\eta$ candidates for (a) the data sample and (b)
    the signal MC sample generated according to the amplitude analysis
    results.}
  \label{dalitz}
\end{figure}

The amplitude analysis requires a sample with good resolution and all
candidates falling within the phase-space boundary. Therefore, a
four-constraint kinematic fit is performed, assuming $D^{-}$ candidates
decaying to one of the tag modes and $D^{+}$ decaying to the signal mode. The
invariant masses of $(\gamma\gamma)_{\eta}$, $(\pi^{+}\pi^{-})_{K_S^0}$, and
$D^{\pm}$ candidates are constrained to their individual known
masses~\cite{PDG}. The Dalitz plot of the $D^+\to K^0_S\pi^+\eta$ candidates for
the data sample is shown in Fig.~\ref{dalitz}(a).

The intermediate-resonance composition is determined by an unbinned 
maximum-likelihood fit. The likelihood function $\mathcal{L}$ is constructed
with a signal probability density function~(PDF), which depends on the momenta
$p$ of the three final-state particles:
$\mathcal{L} = \prod_{k}^{N_{\rm D}}f(p_{j}^{k})$, where $j$ runs over the
three final-state particles, $k$ runs over each data event, $N_{\rm D}$ is the 
number of candidate events in data, and $f$ is the signal PDF. To take the
background into account, its contribution is subtracted from the data by
modifying the negative-likelihood function as
follows~\cite{Langenbruch:2019nwe}:
\begin{eqnarray}\begin{aligned}
    -\ln{\mathcal{L}} = \frac{-N_{\rm D}+wN_{\rm B}} {N_{\rm D}+ w^2N_{\rm B}}
    \bigg[\begin{matrix}\sum\limits_{k}^{N_{\rm D}} \ln f(p_{j}^{k})\end{matrix}-
      \begin{matrix}\sum\limits_{l}^{N_{\rm B}}w \ln f(p_{j}^{l})\end{matrix}\bigg],
    \label{eq:logor1}
\end{aligned}\end{eqnarray}
where $l$ runs over each background event obtained from the inclusive MC
sample, $N_{\rm B}$ is the number of these background events, and the weight
$w=(1-{\rm purity})N_{D}/N_{B}$ is normalized to data based on the signal
purity.

The signal PDF is written as
\begin{eqnarray}\begin{aligned}
  f_{S}(p_{j}) = \frac{\epsilon(p_{j})\left|\mathcal{M}(p_{j})\right|^{2}R_{3}}{\int \epsilon(p_{j})\left|\mathcal{M}(p_{j})\right|^{2}R_{3}\,dp_{j}}\,, \label{signal-PDF}
\end{aligned}\end{eqnarray}
where $\epsilon(p_{j})$ is the detection efficiency and $R_{3}$ is the
standard element of three-body phase space~(see Sec. 49, ``Kinematics,'' in
Ref.~\cite{PDG}). The total amplitude, $\mathcal{M}(p_{j})$, is the coherent
sum of individual amplitudes of intermediate processes,
$\sum \rho_{n}e^{i\phi_{n}}\mathcal{A}_{n}$, where $\rho_{n}$ and $\phi_{n}$
are the magnitude and phase for the amplitude $\mathcal{A}_{n}$ of the $n$th
intermediate resonance, respectively. The amplitude $\mathcal{A}_{n}$ is the
product of the spin factor~\cite{covariant-tensors}, the Blatt-Weisskopf
barriers of the intermediate state and the $D^{+}$
meson~\cite{PhysRevD.83.052001}, and the propagator for the intermediate
resonance. The propagator of $a_0(980)^+$ is parametrized with a Flatt\'e
formula and the parameters are fixed to the values given in
Ref.~\cite{a0980para}. The propagator of $\bar{K}_0^*(1430)^0$ is parametrized
as a relativistic Breit-Wigner~(RBW) function~\cite{RBW} using parameters
obtained in Ref.~\cite{K1430para}. The normalization integral term in the
denominator is handled by MC integration~\cite{BESIII:2022kbq}.

In the fit, the magnitude and phase of the amplitude
$D^{+} \to K_S^0a_0(980)^{+}$ are fixed to 1.0 and 0.0, respectively, while
those of other amplitudes are left floating. Various combinations of amplitudes
for intermediate resonances are tested. The statistical significance of each
amplitude is calculated based on the change of the log-likelihood with and
without this amplitude after taking the change of the degrees of freedom into
account. Two dominant amplitudes with significance greater than 5$\sigma$,
$D^{+} \to K_S^0a_0(980)^{+}$ and $D^{+} \to \bar{K}_0^*(1430)^0\pi^{+}$,
remain as the nominal set. Other amplitudes for possible intermediate
resonances, including $D^{+} \to \bar{K}_1(1270)^0\pi^+$,
$D^{+} \to K^*(1410)^+\eta$, $D^{+} \to K_2^*(1430)^+\eta$,
$D^{+} \to \bar{K}_2^*(1580)^0\pi^+$, $D^{+} \to K_S^0\pi_1(1400)^+$,
$D^{+} \to K_S^0a_0(1450)^+$, $D^{+} \to K_S^0a_2(1320)^+$,
$D^{+} \to K^*(892)^+\eta$,
and $D^{+} \to (K_S^0\pi^+)_{S-{\rm wave}}\eta$~~(using the LASS
parametrization~\cite{KpiLASS}), have no significant contribution.
The contribution of the $n$th intermediate process relative to the total BF is
quantified by a fit fraction~(FF) defined as
\begin{eqnarray}\begin{aligned}
    {\rm FF}_{n} = \frac{\sum^{N_{\rm gen}} \left|\rho_{n}e^{i\phi_{n}}\mathcal{A}_{n}\right|^{2}}{\sum^{N_{\rm gen}} \left|\mathcal{M}\right|^{2}}\,, \label{Fit-Fraction-Definition}
\end{aligned}\end{eqnarray}
where $N_{\rm gen}$ is the number of the generated phase-space MC events. The
statistical uncertainties of FFs are determined by sampling randomly according
to the Gaussian distributions of FF$_{a_0(980)^+}$ and
FF$_{\bar{K}_0^*(1430)^0}$ and to their correlation through the covariance
matrix.

\begin{figure*}[!htbp]
  \centering
  \includegraphics[width=0.3\textwidth]{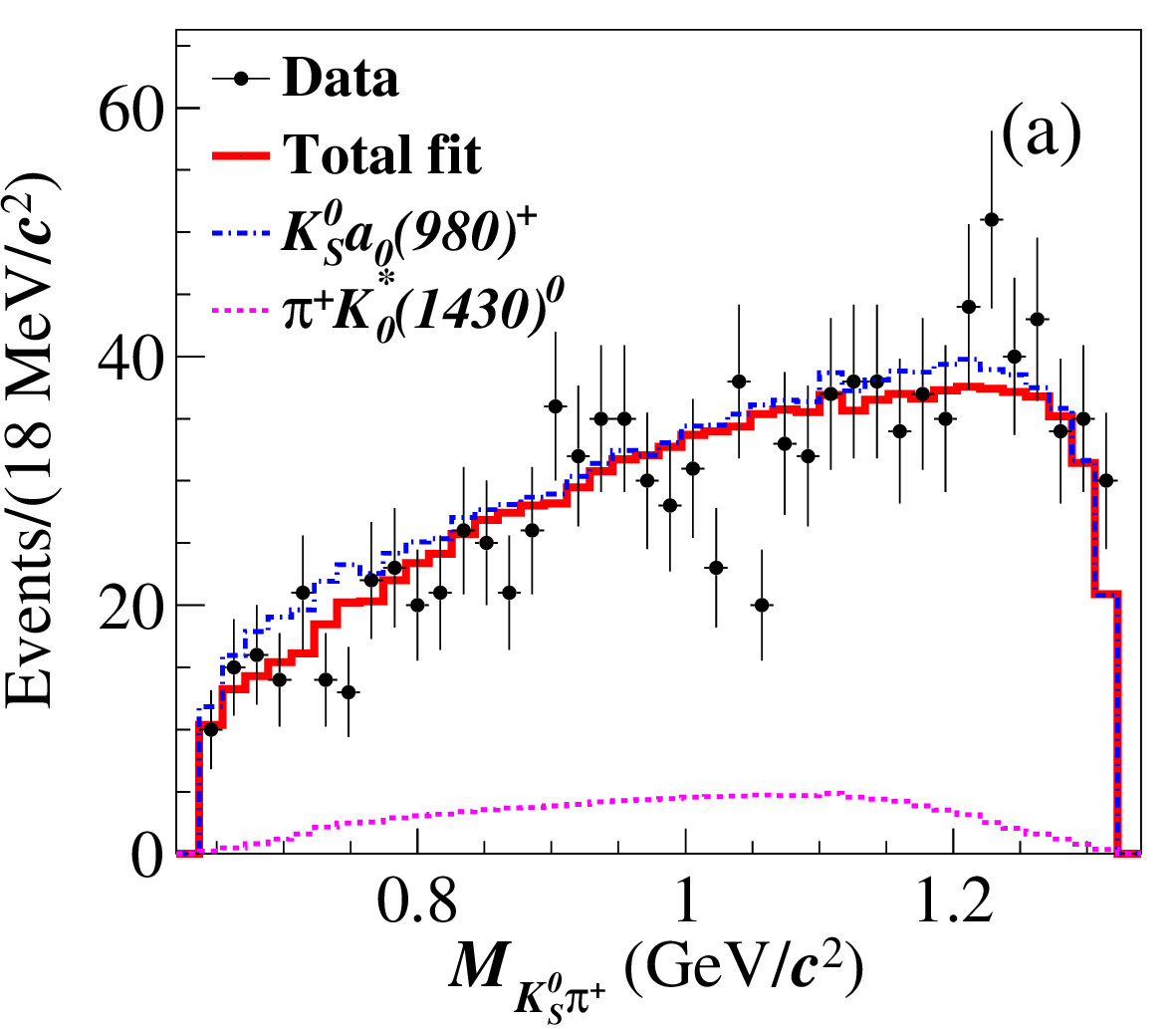}
  \includegraphics[width=0.3\textwidth]{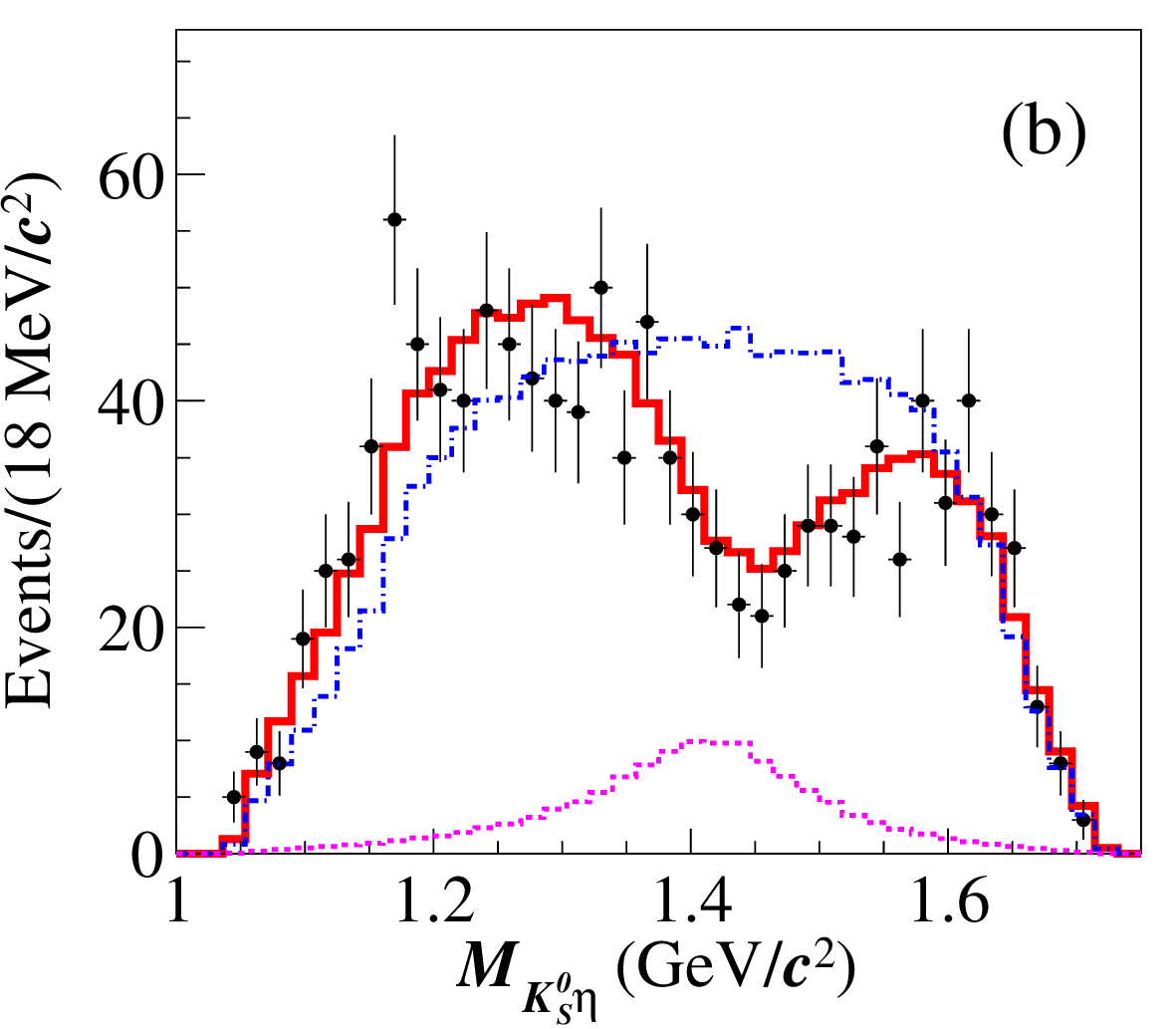}
  \includegraphics[width=0.3\textwidth]{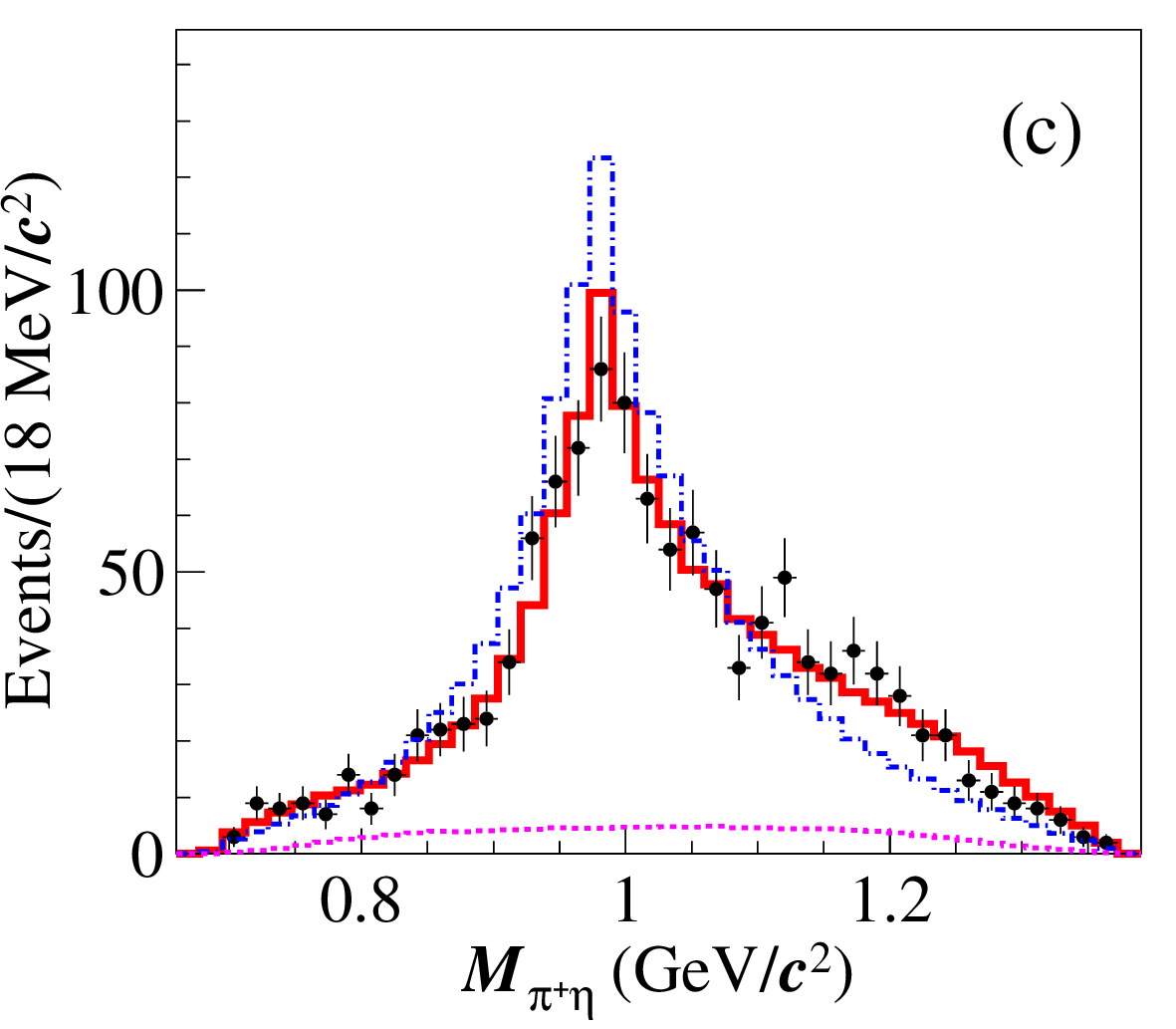}
  \caption{Projections on the invariant masses of (a) $K_S^0\pi^+$, (b)
    $K_S^0\eta$, and (c) $\pi^+\eta$ systems of the nominal fit. The data are
    represented by black dots with the error bars, the fit results by the solid
    red curves with the $K_{S}^{0}a_0(980)^+$ yields by the dashed blue curves
    and the $\bar{K}^*_0(1430)^0\pi^+$ yields by the dashed magenta curves. The
    fit results take into account the relative phases and interferences between
    the two signal amplitudes, making a simple visual addition of the lines of
    the two signal amplitudes insufficient. The impact of the destructive
    interference is clearly visible in (b). Background is not shown as it is
    almost negligible graphically.
  }
  \label{dalitz-projection}
\end{figure*}

The phases, FFs, and statistical significances for the amplitudes are listed in
Table~\ref{fit-result}. The Dalitz plot for the signal MC sample generated
based on the amplitude analysis model is shown in Fig.~\ref{dalitz}(b). The
Dalitz plot projections are shown in Fig.~\ref{dalitz-projection}. The
correlation matrix is provided in Table~\ref{tab:CM}. The systematic
uncertainties of the amplitude analysis, as summarized in
Table~\ref{systematic-uncertainties}, are estimated as described below.
\begin{table*}[htbp]
  \caption{Phases, FFs, and statistical significances for different amplitudes.
    The first uncertainty is statistical, while the second and third
    uncertainties are model and experimental systematic uncertainties,
    respectively. The total of the FFs is not necessarily equal to 100\% due to
    interference effects.}
  \label{fit-result}
  \begin{center}
    \begin{tabular}{lccc}
      \hline
      \hline
      Amplitude                              & Phase $\phi$ (rad)               & FF~(\%)                             &Significance\\
      \hline
      $D^{+} \to K_S^0a_0(980)^+$            & 0.0(fixed)                         & $105.00 \pm 0.94 \pm 1.04 \pm 0.07$ &$>10\sigma$ \\
      $D^{+} \to \bar{K}_0^*(1430)^0\pi^{+}$ & $ 2.58 \pm 0.06 \pm 0.09 \pm 0.01$ & $10.83  \pm 1.50 \pm 1.27 \pm 0.08$ &$>10\sigma$\\
      \hline
      \hline
    \end{tabular}
  \end{center}
\end{table*}
\begin{table}[htbp]
  \renewcommand\arraystretch{1.25}
  \caption{Correlation matrix for the phases and FFs.}
  \label{tab:CM}
  \begin{center}
    \begin{tabular}{l|ccc}
      \hline\hline
      & FF$_{a_0(980)^+}$ & FF$_{\bar{K}_0^*(1430)^0}$ & $\phi_{\bar{K}_0^*(1430)^0}$\\
      \hline
      FF$_{a_0(980)^+}$            & 1.00              & $-0.28$                     & 0.88       \\
      FF$_{\bar{K}_0^*(1430)^0}$   &                   & 1.00                       & 0.20       \\
      $\phi_{\bar{K}_0^*(1430)^0}$ &                   &                            & 1.00        \\
      \hline\hline
    \end{tabular}
  \end{center}
\end{table}

The systematic uncertainty due to the $a_0(980)^+$ line shape is estimated by 
varying the mass and coupling constant of the Flatt\'e propagator by
$\pm 1\sigma$ and a three-channel-coupled Flatt\'e formula which adds the
$\pi\eta^\prime$ according to Ref.~\cite{a0980para}. The systematic uncertainty
caused by the $\bar{K}_0^*(1430)^0$ line shape is estimated by varying the mass
and width of the RBW by $\pm 1\sigma$ according to Ref.~\cite{K1430para}. The
quadratic sum of these two uncertainties is taken to be the amplitude model
systematic uncertainty. The systematic uncertainty due to the effective radius
of the Blatt-Weisskopf barrier~\cite{PhysRevD.83.052001} is determined by
varying the effective radius within the range $[2.0, 4.0]$~GeV for intermediate
resonances and $[4.0, 6.0]$~GeV for $D^+$ mesons. The maximum variations are
taken as the systematic uncertainties. The uncertainties caused by possible but
insignificant intermediate resonances, such as $\bar{K}_1(1270)^0\pi^+$,
$K^*(1410)^+\eta$, $\bar{K}_2^*(1580)^0\pi^+$, $K_S^0a_0(1450)^+$, and
$K_S^0a_2(1320)^+$, are taken to be the differences of the phases and FFs with
and without the intermediate resonances.

To determine the systematic uncertainty related to the background estimation,
the sPlot technique~\cite{Pivk:2004ty} is instead employed on the $M_{\rm BC}$
variable to eliminate the combinatorial background. An amplitude analysis,
after applying sWeights, is performed, and deviations from the nominal results
are assigned as the systematic uncertainties. The uncertainty associated with
the detector acceptance difference between MC samples and data is determined by
reweighting $\epsilon(p_{j})$ in Eq.~\ref{signal-PDF} with different
reconstruction efficiencies of $K_S^0$, $\pi^+$, and $\eta$ according to their
uncertainties. The changes of the fit results are taken as the systematic
uncertainties. In addition, with the amplitude analysis results obtained in
this work, 300 signal MC samples are generated with the same size as data. The
pull value is calculated by
$V_{\rm pull}=(V_{\rm fit}-V_{\rm input})/\sigma_{\rm fit}$, where
$V_{\rm input}$ is the input value in the generator and $V_{\rm fit}$ and
$\sigma_{\rm fit}$ are the output value and corresponding statistical
uncertainty, respectively. The resulting pull distributions are consistent
with the standardized normal distributions, showing no significant fit bias.

\begin{table}[htbp]
  \caption{Systematic uncertainties on the phases and FFs for different
    amplitudes in units of statistical uncertainties. The sources are
    categorized as: Model - (I) Amplitude model, (II) Effect radius, and (III)
    Insignificant resonances; Experimental~(Exp.) - (IV) Background and (V)
    Detector acceptance.
  }
  \centering
  \begin{tabular}{lcccccccc}
    \hline
    \hline
    \multirow{2}{*}{Amplitude}&\     &\multicolumn{3}{c}{Model}&\multicolumn{2}{c}{Exp.}&\\
    &\      &I    &II   &III  &IV   &V    &Total\\
    \hline
    $D^+ \to K_S^0a_0(980)^+$
    &FF     &0.32 &0.01 &1.06 &0.10 &0.04 &1.11\\	
    \hline
    \multirow{2}{*}{$D^+ \to \pi^+\bar{K}_0^*(1430)^0$}
    &$\phi$ &0.47 &0.01 &1.33 &0.13 &0.08 &1.42\\
    &FF     &0.12 &0.01 &0.84 &0.04 &0.05 &0.85\\	
    \hline
    \hline
  \end{tabular}
  \label{systematic-uncertainties}
\end{table}

The BF of $D^+ \to K_S^0\pi^+\eta$ was previously measured in
Ref.~\cite{PRL124.241803}. We update this BF to be
$(1.27\pm0.04_{\rm stat.}\pm0.03_{\rm syst.})\%$ with a signal MC sample
generated based on the amplitude analysis model, which provides a more precise
estimation of the detection efficiency. The uncertainty associated with the
amplitude analysis model, 0.7\%, is estimated by varying the model parameters
based on their error matrix.

In summary, we perform an amplitude analysis of $D^+ \to K_S^0\pi^+\eta$ for
the first time and report the observation of $D^+\to K_S^0 a_0(980)^+$. The
interferences between intermediate resonances are fully considered and a
($15.83\pm1.53_{\rm stat.}\pm1.65_{\rm syst.})\%$ destructive interference is
observed between the $D^+\to K_S^0 a_0(980)^+$ and
$D^+\to \bar{K}_0^*(1430)^0\pi^+$ amplitudes. The amplitude analysis results
are listed in Table~\ref{fit-result}. With a detection efficiency obtained with
MC samples according to the amplitude analysis model, we obtain
$\mathcal{B}(D^+ \to K_S^0\pi^+\eta) = (1.27\pm0.04_{\rm stat.}\pm0.03_{\rm syst.})\%$.
This work uses the same datasets as the previous measurement of
$(1.31\pm0.05)\%$ did~\cite{PDG, PRL124.241803}, whereas the small difference
arises from the change of the signal model from a modified data-driven
generator BODY3~\cite{BODY3} to the amplitude analysis model, which is used to
estimate the detection efficiency.

Using the FFs listed in Table~\ref{fit-result}, the BFs for the intermediate
processes are calculated as
$\mathcal{B}_i={\rm FF}_i\times \mathcal{B}(D^+ \to K_S^0\pi^+\eta)$ to be
$\mathcal{B}(D^+\to K_S^0a_0(980)^+, a_0(980)^+\to \pi^+\eta) = (1.33 \pm 0.05_{\rm stat.}\pm 0.04_{\rm syst.})\%$
and
$\mathcal{B}(D^+\to \bar{K}_0^*(1430)^0\pi^+, \bar{K}_0^*(1430)^0\to K_S^0\eta) = (0.14 \pm 0.02_{\rm stat.} \pm 0.02_{\rm syst.})\%$.
Theoretical studies insistently require experimental measurements of
$D^+ \to K_S^0a_0(980)^+$, as its observation can provide sensitive
constraints in the extraction of contributions from internal $W$-emission
diagrams of $D \to SP$~\cite{Cheng:2002ai, Cheng:2010vk, Cheng:2022vbw}. If
these measured BFs cannot be well described by the diagrammatic approach it
would indicate that significant final-state interactions must be
involved~\cite{Achasov:2021dvt}, which will provide critical information on the
role of $a_0(980)$ in charmed meson decays and the nature of $a_0(980)$. In
addition, the BF of $D^+\to \bar{K}_0^*(1430)^0\pi^+$ obtained in this work,
$(3.26 \pm 0.47_{\rm stat.} \pm 0.47_{\rm syst.}\,^{+1.02}_{-1.29K_0^*})\%$\footnote{The third uncertainty comes from the BF of $\bar{K}_0^*(1430)^0\to K^0\eta$, $(8.6^{+2.7}_{-3.4})\%$~\cite{PDG}.},
is consistent with that derived from $D^+\to K^-\pi^+\pi^+$ and
$D^+\to K_S^0\pi^+\pi^0$ by the PDG,
$(2.02 \pm 0.10_{\rm stat.} \pm 0.22_{\rm syst.})\%$~\cite{PDG}.
In the near future, we plan to use 20~fb$^{-1}$ data to conduct a more
precise investigation into the production of $a_0(980)$ and $f_0(980)$ in
$D$-meson decays~\cite{BESIII:2020nme, Ke:2023qzc, Li:2021iwf}.

\acknowledgments
The BESIII Collaboration thanks the staff of BEPCII and the IHEP computing center for their strong support. The authors thank Prof. Yu-Kuo Hsiao for helpful discussions. This work is supported in part by National Key R\&D Program of China under Contracts Nos. 2020YFA0406400, 2020YFA0406300; National Natural Science Foundation of China (NSFC) under Contracts Nos. 11635010, 11735014, 11835012, 11875054, 11935015, 11935016, 11935018, 11961141012, 12022510, 12025502, 12035009, 12035013, 12061131003, 12192260, 12192261, 12192262, 12192263, 12192264, 12192265, 12221005, 12225509, 12235017, 12205384; the Chinese Academy of Sciences (CAS) Large-Scale Scientific Facility Program; the CAS Center for Excellence in Particle Physics (CCEPP); Joint Large-Scale Scientific Facility Funds of the NSFC and CAS under Contract Nos. U2032104, U1832207; CAS Key Research Program of Frontier Sciences under Contracts Nos. QYZDJ-SSW-SLH003, QYZDJ-SSW-SLH040; 100 Talents Program of CAS; The Institute of Nuclear and Particle Physics (INPAC) and Shanghai Key Laboratory for Particle Physics and Cosmology; ERC under Contract No. 758462; European Union's Horizon 2020 research and innovation programme under Marie Sklodowska-Curie grant agreement under Contract No. 894790; German Research Foundation DFG under Contracts Nos. 443159800, 455635585, Collaborative Research Center CRC 1044, FOR5327, GRK 2149; Istituto Nazionale di Fisica Nucleare, Italy; Ministry of Development of Turkey under Contract No. DPT2006K-120470; National Research Foundation of Korea under Contract No. NRF-2022R1A2C1092335; National Science and Technology fund of Mongolia; National Science Research and Innovation Fund (NSRF) via the Program Management Unit for Human Resources \& Institutional Development, Research and Innovation of Thailand under Contract No. B16F640076; Polish National Science Centre under Contract No. 2019/35/O/ST2/02907; The Swedish Research Council; U. S. Department of Energy under Contract No. DE-FG02-05ER41374

\end{document}